# Development of grid-based and PINN solvers for electron kinetics in collisional non-thermal plasmas

**Vladimir Kolobov[1*] and Lucius Schoenbaum[1]**

[1] University of Alabama in Huntsville, Huntsville, AL, USA

[*]E-mail: vik0001@uah.edu

**Abstract.** We compare traditional finite volume and Physics Informed Neural Network (PINN) solvers for elliptic (Poisson), hyperbolic (advection), and parabolic (diffusion) equations in 2d settings. We describe the challenges of using traditional and PINN solvers for electron kinetic equations in collisional plasmas. The advantages and drawbacks of PINNs over state-of-the-art traditional solvers are discussed. We also consider angular moments in spherical velocity space and the potential use of ML algorithms for reduced kinetic models in the coordinate-energy phase space based on adaptive closure relations.

## 1. Introduction

Inspired by the explosive growth in publications devoted to Physics Informed Neural Networks (PINNs), we have explored their potential to solve kinetic equations for electrons in collisional plasma. We have previously developed grid-based solvers for the kinetic equations that are attractive for hybrid kinetic-fluid plasma solvers [1]. They provide a noise-free alternative to the commonly used particle-based (PIC) methods. Expected potential challenges are due to the non-locality of the collision operators in velocity space and the multi-scale nature of the kinetic equations (the presence of a diffusion kinetic scale between the fully kinetic and fluid limits). An integral operator in velocity space describes the large-angle elastic scattering of electrons on atoms (rather than the Fokker Planck differential model for small-angle scattering). Inelastic collisions associated with an energy loss quantum larger than electron temperature are described by a non-local term with a shifted argument (in energy) rather than a continuum loss model associated with quasi-elastic collisions. The dominance of elastic scattering over inelastic collisions with energy loss results in an intermediate (diffusion) time scale in a phase space of reduced dimensions [2]. The closure of angular moments is a potential use of ML [3].

We first compare PINNs with Finite Volume (FV) solvers with adaptive Cartesian mesh (ACM) for elliptic (Poisson), hyperbolic (advection), and parabolic (diffusion) solvers. Then, we explore PINNs to solve kinetic equations with non-local scattering operators and inelastic collisions. We discuss the advantages and drawbacks of PINNs to state-of-the-art traditional solvers using adaptive mesh in phase space.

## 2. Kinetic equations for collisional plasmas

*2.1 Kinetic equation for electrons*

Using spherical coordinates in velocity space, the kinetic equation for the electron velocity distribution function (EVDF) $f$ can be written in the form [4]:

$$\frac{\partial f}{\partial t} + v\mu\frac{\partial f}{\partial x} - \frac{eE}{m}\left(\mu\frac{\partial f}{\partial v} + \frac{1-\mu^2}{v}\frac{\partial f}{\partial \mu}\right) = S\,. \quad (1)$$

Here $\mu = cos\,\theta$, is the cosine of the pitch angle $\theta$ between axial direction, $v > 0$ is the velocity amplitude, $e$ and $m$ are the electron charge and mass, and $S$ is a collision term.

Collisions of electrons with atoms can be divided into three types. The first type is elastic scattering, described by a Lorentz model [5]. For isotropic scattering:

$$S_{el} = \nu(v)\left[\int_{-1}^{1} f(\mu')d\mu' - f(v,\mu)\right] = \nu(v)\big(f_0(v) - f(v,\mu)\big)\,, \quad (2)$$

where $\nu(v)$ is the elastic collision frequency. The operator (2) conserves the electron energy and the number of electrons. Small-angle elastic scattering (grazing collision limit) is described by the Fokker-Planck-Lorentz model [6]:

$$S_{el} = \nu(v)\frac{\partial}{\partial \mu}\left((1-\mu^2)\frac{\partial f}{\partial \mu}\right). \quad (3)$$

The collision operators (2) and (3) modify only the direction of particle motion and conserve the modulus of the particle velocity or its kinetic energy. They are also used in the radiation transport of photons, which move at a constant speed (constant $v$).

The second type of electron-atom collision is an inelastic collision associated with the excitation of atoms. For example, the excitation of the first atomic level with energy $\varepsilon_1$ is described by [7]:

$$S_{exc} = \nu_{ex}(v^*)f_0(v^*)\frac{v^*}{v} - \nu_{ex}(v)f(v)\,, \quad (4)$$

where $\nu_{ex}(v)$ is the corresponding inelastic collision frequency, and $v^* = \sqrt{v^2 + 2\varepsilon_1/m}$. This operator conserves the number of particles,

$$n_e = \int_0^\infty f_0(v)\,v^2 dv\,. \quad (5)$$

*2.2 Angular moments and quasi-diffusion models*

A common approach for reducing the dimensionality of transport problems is using angular moments, $f_0 = \int_{-1}^{1} f(\mu)d\mu$, $f_1 = \int_{-1}^{1} \mu f(\mu)d\mu$, $f_2 = \int_{-1}^{1} \mu^2 f(\mu)d\mu$ , etc. We obtain from (1) two coupled equations for $f_0(x,v)$ and $f_1(x,v)$:

$$\frac{\partial f_0}{\partial t} + v\frac{\partial f_1}{\partial x} + \frac{eE}{mv^2}\frac{\partial}{\partial v}(v^2 f_1) = S_0\,. \quad (6)$$

$$\frac{\partial f_1}{\partial t} + v\frac{\partial \chi f_0}{\partial x} + \frac{eE}{m}\frac{\partial \chi f_0}{\partial v} + \frac{eE}{mv}(3\chi - 1)f_0 = -\nu f_1\,, \quad (7)$$

where $\nu$ is the transport collision frequency and $\chi = \frac{f_2}{f_0}$ is an Eddington factor [3]. For $\chi = 1/3$ , one obtains a Lorentz model, $f = f_0 + \mu f_1$ , with the equation for $f_1$:

$$\frac{\partial f_1}{\partial t} + \frac{v}{3}\frac{\partial f_0}{\partial x} + \frac{eE}{3m}\frac{\partial f_0}{\partial v} = -\nu f_1 \,. \tag{8}$$

The case of $\chi = 1$ corresponds to a mono-directed electron beam, when $f = f_0(x, v)\delta(\mu - 1)$.

Returning to the general case. Neglecting the time derivative in (7) and using the electron kinetic energy $u$ expressed in eV as an independent variable, we obtain the system [8]:

$$v\frac{\partial f_0}{\partial t} + \left(\frac{\partial uJ}{\partial x} + E\frac{\partial uJ}{\partial u}\right) = S_0 \,, \tag{9}$$

$$J = -\frac{1}{N\sigma_{tr}}E\left(\frac{\partial \chi f_0}{\partial u} + \frac{1}{2u}(3\chi - 1)f_0\right) - \frac{1}{N\sigma_{tr}}\frac{\partial \chi f_0}{\partial x} \quad . \tag{10}$$

This system describes quasi-diffusion in phase space $(x, u)$ with a broader applicability range [9] than the two-term spherical harmonics expansion (SHE) described below.

**Two stream model**

The half-moment model, aka two-stream approximation or Shockley model in semiconductor transport, uses two velocity directions [4,10]:

$$f(x, v, \mu) = f_+(x, v)\,\delta(\mu - 1) - f_-(x, v)\,\delta(\mu + 1) \,. \tag{11}$$

For isotropic scattering, we obtain [10]:

$$\frac{\partial f}{\partial t} + v\frac{\partial f}{\partial x} + \frac{eE}{m}\frac{\partial f}{\partial v} = \frac{\nu}{2}\big(f(-v) - f(v)\big) \,. \tag{12}$$

The particle number density in the two-stream model is defined as:

$$n_e = \int_{-\infty}^{\infty} f(v)\, dv \quad . \tag{13}$$

**Lorentz model**

The two-term SHE results in two coupled equations for $f_0(x, v)$ and $f_1(x, v)$ , which evolve on different time scales [8]:

$$\frac{\partial f_0}{\partial t} + \frac{v}{3}\nabla \cdot \boldsymbol{f}_1 - \frac{1}{3v^2}\frac{\partial}{\partial v}\left(v^2 \frac{e\boldsymbol{E}}{m} \cdot \boldsymbol{f}_1\right) = S_0 \,, \tag{14}$$

$$\frac{\partial \boldsymbol{f}_1}{\partial t} + \nu \boldsymbol{f}_1 = -v\nabla f_0 + \frac{e\boldsymbol{E}}{m}\frac{\partial f_0}{\partial v} . \tag{15}$$

As the characteristic time scale of the time-variation $f_0$ is larger than $\nu^{-1}$, a diffusion-type Fokker-Planck (FP) kinetic equation can be obtained for $f_0(x, v)$:

$$\frac{\partial f_0}{\partial t} - \left(\frac{\partial}{\partial x} - E\frac{1}{\sqrt{u}}\frac{\partial}{\partial u}\sqrt{u}\right) \cdot D_r\left(\frac{\partial}{\partial x} - E\frac{\partial}{\partial u}\right) f_0 = S_0 \,, \tag{16}$$

where $D_r = v^2/(3\nu)$ is the spatial diffusion coefficient. This equation has been used to analyze the axial and radial structure of DC discharges and plasma stratification [11]. For weakly ionized plasmas, the gas pressure $p$ defines the amplitude of electron collision frequencies, which are proportional to $p$.

## 3. Grid-based and PINN solvers for PDEs

### *3.1 Elliptic equations*

*PINN predictivity* is the term we have used to describe the ability of PINNs to both interpolate within their training domain and extrapolate to regions outside their training domain. These abilities may be viewed as being a consequence of the neural network's theoretical domain having no true limit, i.e., a neural network with input dimension $n$ may be evaluated at any point in $n$-space up to the scalar hardware limit, together with the observation that training (gradient descent optimizer iterations) affects the neural network output on the entire theoretical domain. As a rule, the neural network's convergence on the training domain degrades gradually, not suddenly, along with the distance from the training domain. The accuracy of extrapolated values can be appreciable, as indicated in **Figure 1**, which demonstrates PINN predictivity using a solution of the two-dimensional Poisson equation in a unit cube with Dirichlet boundary conditions $(f(x=1) = 1 - y^2,\ f(y=1) = x^2 - 1,\ f(y=0) = x^2,\ f(x=0) = -y^2)$. Other authors have observed and commented on similar phenomena [12].

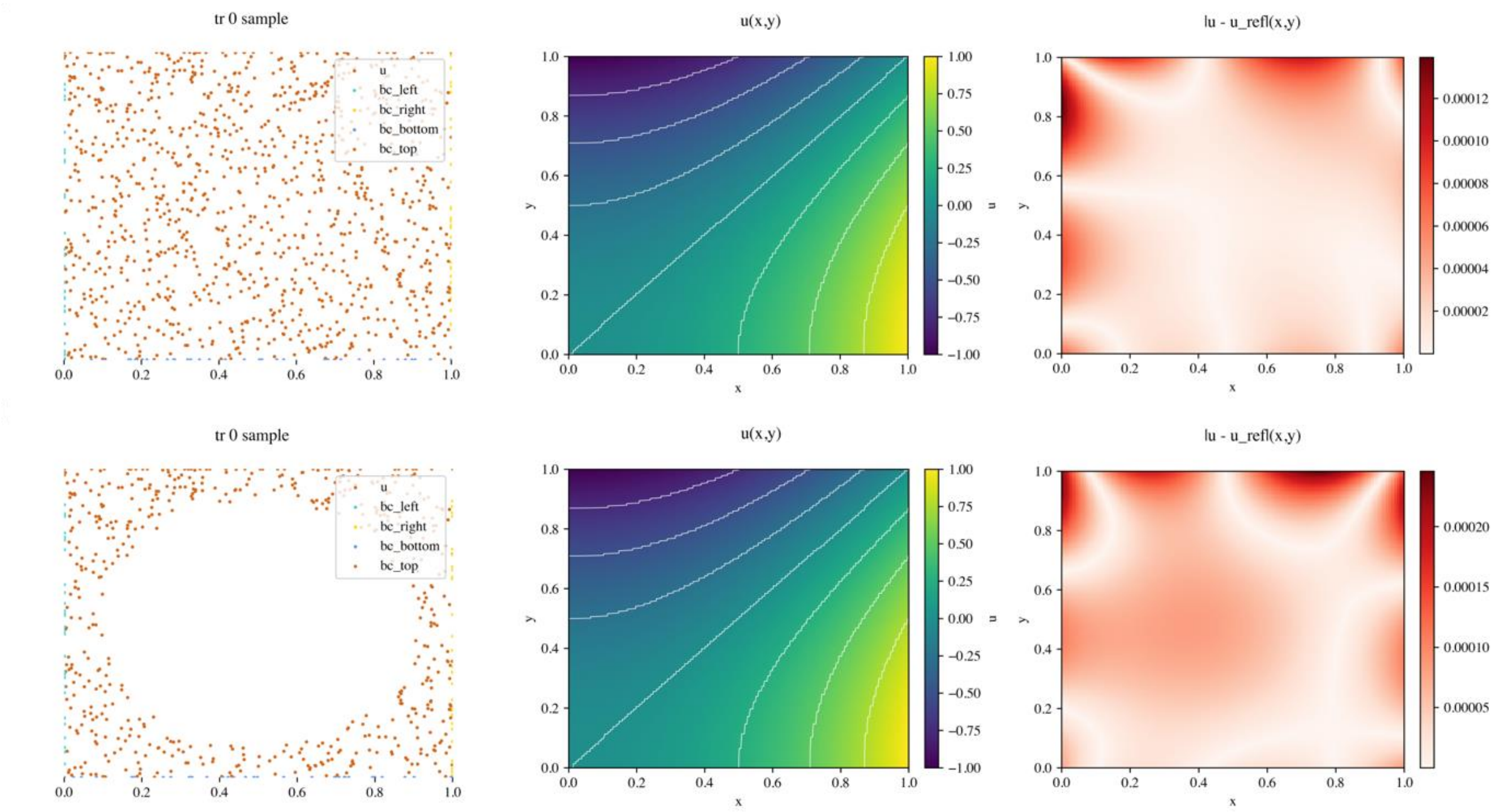

**Figure 1.** Illustration of PINN Predictivity for a Poisson equation with Dirichlet boundary conditions. Unlike traditional solvers, a PINN can effectively interpolate in regions outside its training domain. (Left) sample sets for two cases. (Middle) solution plots for the same two cases. (Right) the $L^1$ error compared to the exact solution $f(x, y) = x^2 - y^2$. For each case, training stopped after the training error fell below $10^{-7}$ (the rightmost plot shows the validation error, not the training error).

Whereas traditional FV solvers find values using information at neighboring cells, PINN optimization considers information at values much further away, as well as values of higher order, that is, derivatives obtained using automatic differentiation (AD). This allows a PINN to produce a smooth interpolation between collocation points even in regions of low density and extrapolate values potentially very far outside of its domain.

*3.2 Hyperbolic equations*

Our investigations found that properly tuned PINNs perform well on 1D-3D advection problems. An example shown in **Figure 2** illustrates a PINN solution for a simple 1D advection problem with periodic boundary conditions [13]:

$$u_t + \beta u_x = 0$$
$$u(x_{min}, t) = u(x_{max}, t)$$
$$u(x, 0) = sin(\pi x)$$

When β is large, the optimizer can fail to make progress toward the proper solution, and instead, the optimizer opts for the best solution it can find, namely, $u(x,t) = 0$, a divergent solution known as a zero failure [13]. Here, we used a small value of β to demonstrate moving time windows (PINN time stepping) and PINN predictivity in the context of time-dependent training.

Moving time windows (PINN time stepping), as shown in the example of Figure 2, is the strategy of breaking up the time axis $[t_{initial}, t_{final}]$ into $N$ subdomains $t_{initial} - t_0 < t_1 < \cdots < t_i < \cdots < t_N = t_{tfinal}$ and solving the PINN separately on each subdomain $[t_i, t_{i+1}]$. In a time-marching problem, moving from $t = t_i$ to $t = t_{i+1}$ can be achieved by evaluating the model after the $i$th timestep on the time slice $t = t_{i+1}$, obtaining a dataset $\mathcal{D}_i$, and updating the initial conditions $IC \leftarrow \mathcal{D}_i$ to reflect the new base values. This way, the PINN can naturally advance in steps, like a traditional FV solver using discrete time steps. The most significant differences are (1) the potential difference in magnitude of the delta between the timesteps of a traditional solver and PINN and (2) the continuous nature of the solution obtained by the PINN through each of its timesteps compared to the discrete solution of the traditional solver. This indicates how, for instance, the PINN's zero failure and other failures due to causality are broadly like a grid-solver CFL violation. While the PINN solver overcomes the strict CFL timestep limitation of the explicit FV solver, the greater freedom (in selecting the timestep) comes at a much higher computational cost for each timestep, and this cost (and risk of failure to converge) generally increases along with the timestep size. Although the resulting trade-off for time-stepping PINNs can be optimized for the lowest computational cost, we expect the resulting computational cost for PINN training to still be higher than that for grid-based traditional solvers, cf. e.g., Ref. [14].

**Figure 2** also shows how PINN predictivity can extend beyond the convex hull of the training domain. This allows the PINN to converge to an approximate solution in the following (and previous) timestep. In other words, it can "anticipate" features of the next step. Other tests (not shown) indicate that this property holds even when the training domain is very thin (even as thin as a single time slice). This observation can be applied to PINN training strategies needed for kinetic models (see section 4).

The second timestep requires fewer iterations to converge than the first due to the "burn-in" effect of already solving for the earlier timestep on the same model parameters. The model output in the first timestep (top row) is still working to converge after 2100 iterations (middle column), while in the second timestep (bottom row), the model output has converged well after 2100 iterations (right column). This can be regarded as a form of transfer learning [15].

Finally, we note that in obtaining this solution, we find it helpful to rely on the time-adaptive weighting of the simple and efficient type proposed in Ref. [16].

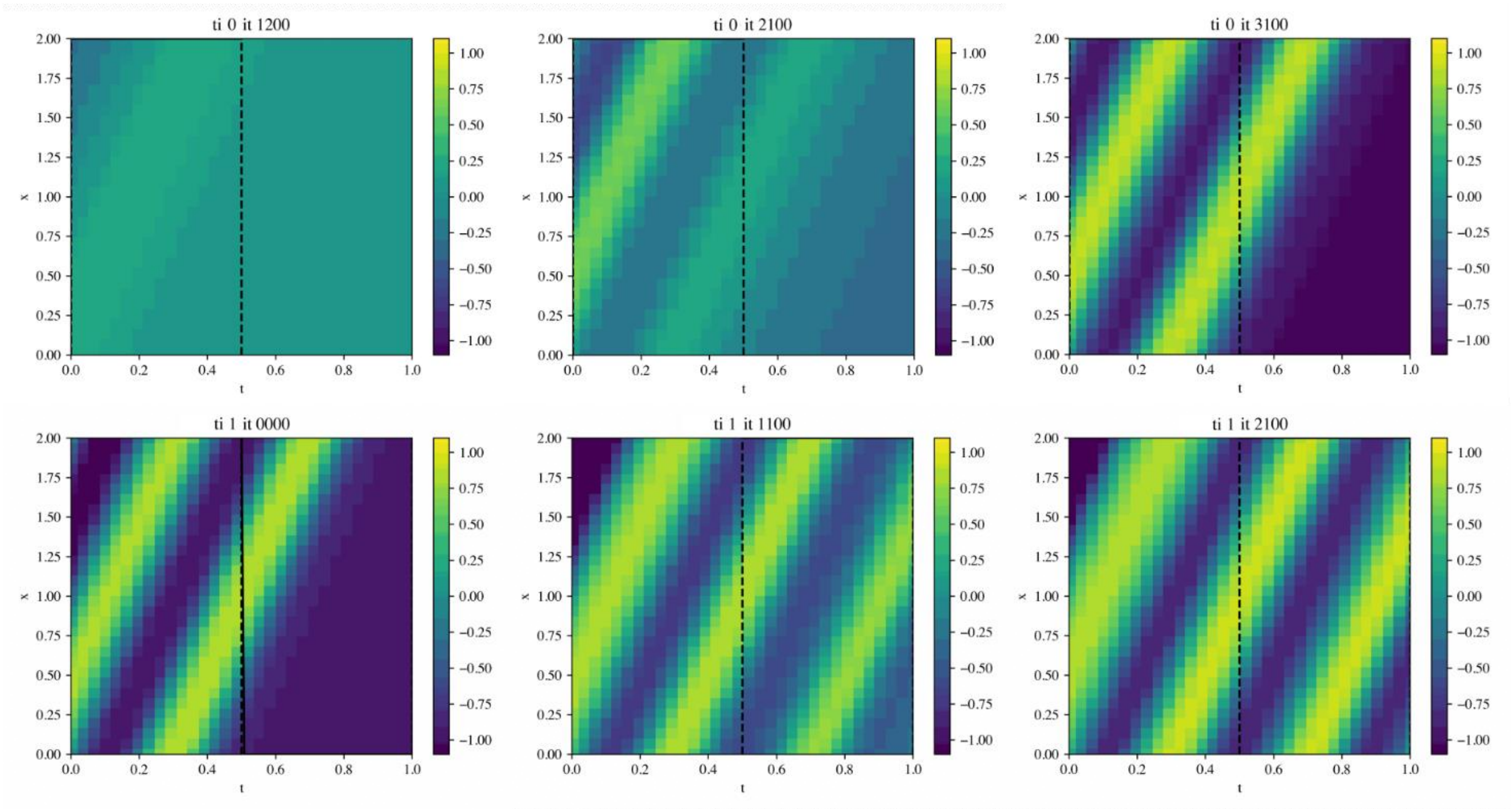

**Figure 2.** Time-stepping PINN solver during training, illustrating PINN predictivity. Unlike for FV solvers, PINN timesteps can occur at the problem scale. In this case, there are two timesteps (indexed by variable $ti = 0, 1$) divided by a vertical dashed line, with time $t$ shown on the $x$-axis. Three snapshots during training of the first ($t = 0$ to 0.5) timestep are displayed on the top row, and similar snapshots are shown for the second ($t = 0.5$ to 1) timestep on the bottom row (indexed by training iteration variable $it$). In the top row, the PINN is trained only on the left rectangular subdomain (i.e., the first timestep). Still, by the end of training (top right), the traveling wave extends well into the right rectangular subdomain (i.e., the next timestep), illustrating another predictivity effect. The periodicity constraint $u(x_{max}, t) = u(x_{min}, t)$ is violated in the second timestep until training commences there (bottom left). When this finishes (bottom right), the second timestep is fully solved, while the solution in the first timestep (being no longer targeted) begins to degrade but still exhibits qualitative features of its solution.

*3.3 Parabolic equations*

**Figure 3** compares FV solver results with PINN results for a 3D diffusion over a sphere. A similar problem for a 2d diffusion over a ring was described in [11]. For the PINN simulation, a fully connected architecture with seven layers and 512 neurons per layer was trained using the Adam optimizer, with adaptive activations and tf-exponential learning rate $\gamma = 0.9$. The Modulus PINN solver package (formerly called SimNet) [13] was used for this simulation.

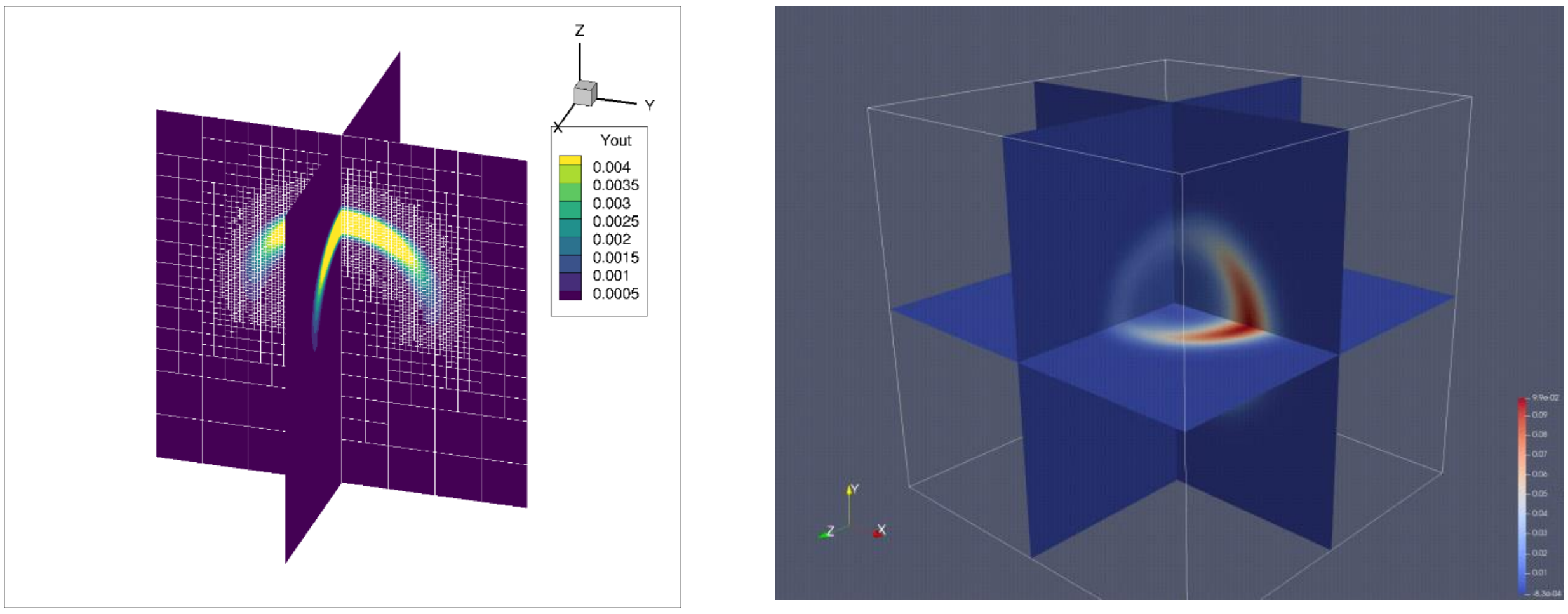

**Figure 3.** Comparison of FV and PINN results for diffusion over a sphere.

With PINN training on marching problems, maintaining accuracy with increasing time or training over long time intervals can be challenging. Temporal loss weighting improved performance somewhat. However, we were motivated to pursue a more robust solution for the time-marching PINN problems described in the previous section.

## 4. Kinetic solvers

### *4.1 Two-stream model (1d1v)*

**Figure 4** compares FV and PINN results for the two-stream model, with $E = 0.4$ and elastic scattering only. Particles are injected at the cathode $(x = 0)$ with positive speed $v$. White lines show characteristics (particle trajectories) in $(x, v)$ phase space. For the PINN solution, 60 iterations of high-intensity LBFGS training were used, with an exponentially decreasing weight on the initial conditions (decreasing from 100 to 1 as training progresses). Additional collocation points were stacked near the origin $(x = 0, v = 0)$ to improve the PINN solution there, suggesting that a more refined adaptive algorithm for collocation points could be used to avoid problem-dependent adjustments.

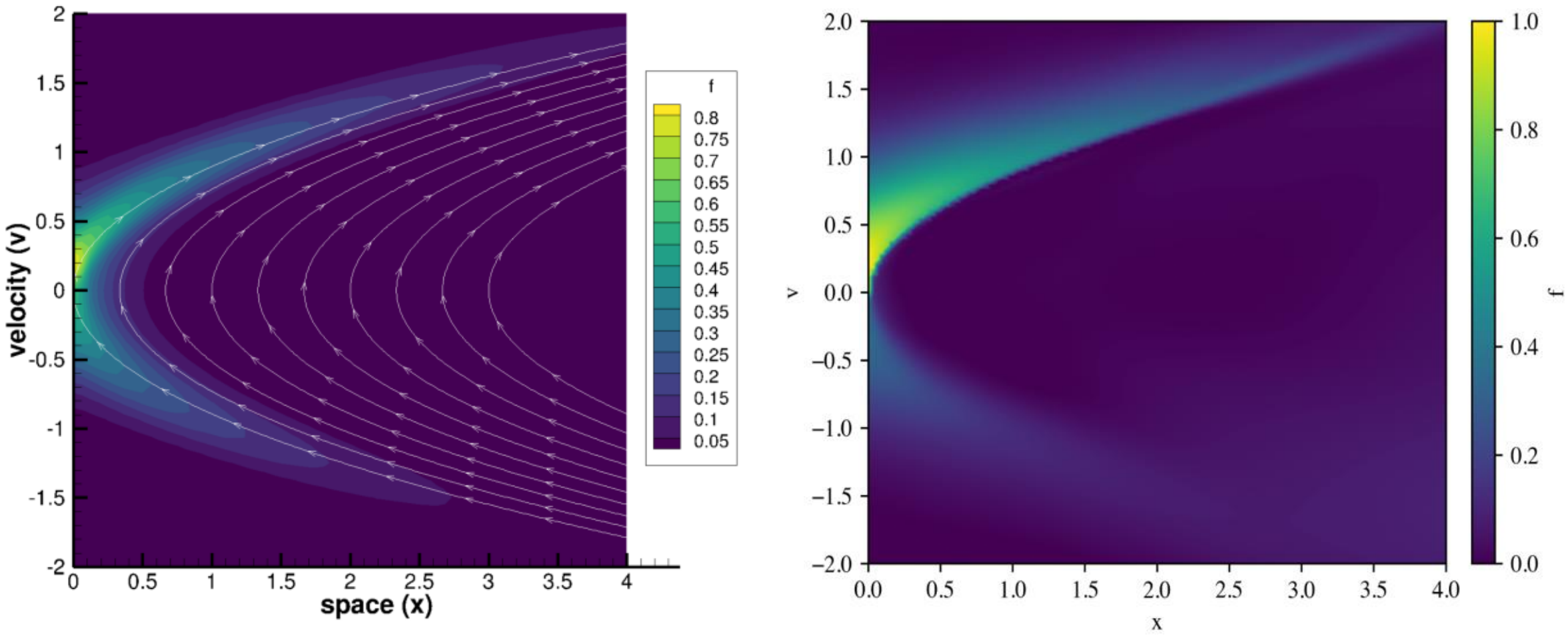

**Figure 4.** Comparison of PINN (right) with FV solver (left) for the two-stream model

A comparison of low vs. high-intensity LBFGS optimizer and the effect of spatial resolution indicated that tuning/configuring PINNs is problem-dependent. There is a lack of general methods for the PINN's underlying multi-objective optimization problem.

*4.2 Radiation transport (1d1$\mu$)*

For $|v|=1$, Eq. (1) with the Lorentz scattering model (2) or the Fokker-Planck-Lorentz model (3) becomes two-dimensional. Consider electrons injected at the left boundary $x=0$ with an angular distribution $g(\mu)$ and absorbed at the right boundary $x=L$. The boundary conditions are:

$$f(0,\mu>0)=g(\mu) \qquad f(L,\mu<0)=0 \tag{17}$$

The angular distribution of the particles formed within the slab depends on the Knudsen number $Kn=1/(\nu L)$. We have compared results for the isotropic scattering (Lorentz) and small angle scattering (FPL) models, as was done in [6] for the angular distribution of injected particles in the form:

$$g(\mu)=\frac{1}{\pi\Delta}exp\left\{-\frac{\theta^2}{\Delta}\right\} \tag{18}$$

**Figure 5** shows the particle density profile and the flux density in a steady state for $\Delta=0.1$. The FV code approximately conserves the flux density, and the results are in good agreement with the results of [6].

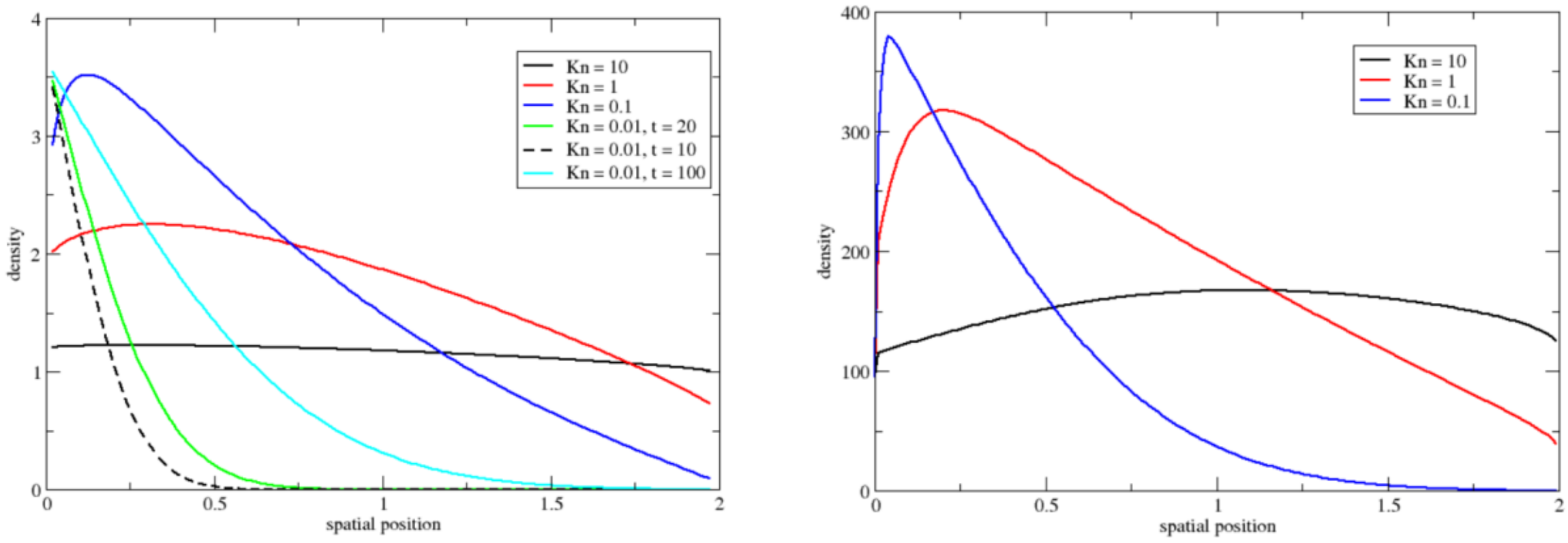

**Figure 5:** Particle density for isotropic scattering (left) and small angle scattering (right) at different $Kn$ numbers computed with FV solver.

**Figure 6** compares VDFs for isotropic scattering and small angle scattering.

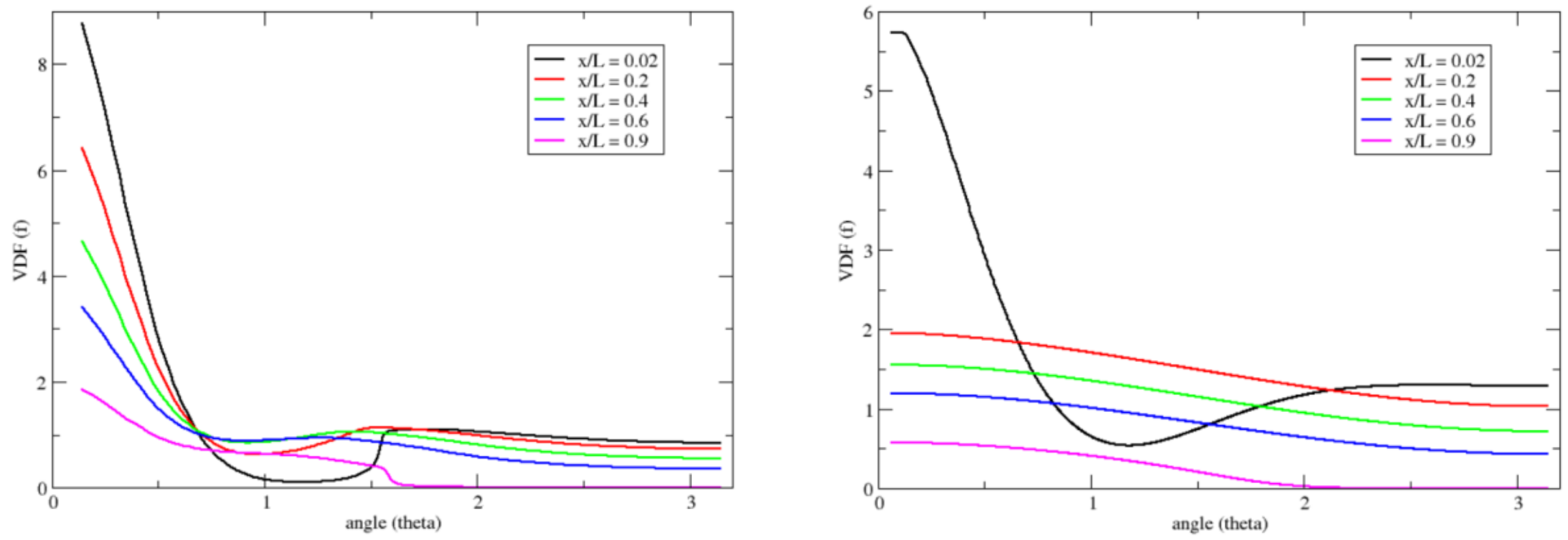

**Figure 6:** Steady state VDFs for isotropic scattering (left) and small angle scattering (right) at $Kn = 1$.

**Figure 7** shows PINN results for a similar 1d1μ case to illustrate PINN's ability to compute the integral scattering term for radiation transport. The only significant error between the FV solver and PINNs is around the point $(x, v) = (0,0)$. Using time slices allows the integral scattering term to be computed only $N$ times (where $N$ is the number of time slices) instead of computing (in the worst case) once for each point of the sample size. For each time slice, integrals are calculated for each point on a mesh and values at mesh points are used to interpolate for final values at collocation points. Because the values are kept in computer memory, the integral scattering term can be recalculated only for a subset of optimizer iterations. This results in a very efficient, flexible computation without significantly impacting the PINN solution.

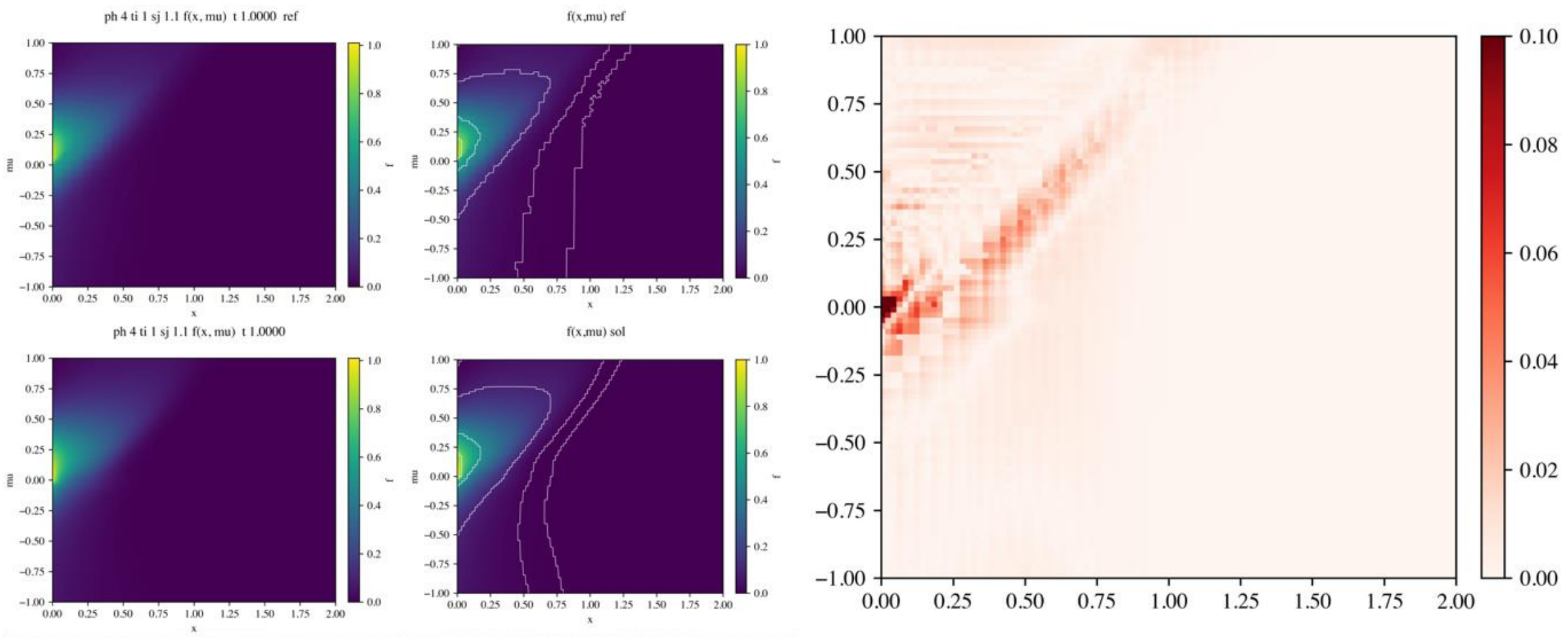

**Figure 7:** PINN results for the 1d1μ isotropic scattering problem at Knudsen number $Kn$ = 0.5. (Top row, left) reference solution (from an FV solver) with and without contour lines. (Bottom row, left) the solution from time-stepping PINN at the final time slice. The training uses two steps with five slices each, separated by four gaps ($\Delta t = 0.125, t_{final} = 1$). (Right) $L^1$ validation error comparing the PINN to the reference solution (both shown at left) showing a median error value of 0.00034 (mean error value 0.0029). The maximum error 0.35 is isolated in the region nearby (0,0), where the PINN solution smoothly models the wavefront near the sharp cutoff of the inflow region $x = 0, \mu > 0$.

*4.3 PINN training procedure (1d1μ case)*

As we did in section 3.2, we set up the PINN for time-stepping, and we split the time domain $t = 0$ to 1 into two timesteps. We targeted five slices per timestep on which to calculate the integral $f_0$ for isotropic scattering, see Equation (2). Because of PINN predictivity effects, increasing the number of slices does not significantly improve accuracy. Because the source term depends on the neural network's output (via evaluation of $f$ and of $f_0$), during the first preliminary phase, we trained on the PDE with the source term set equal to zero. During this phase, we used a high-intensity LBFGS optimizer and trained for 8 iterations to quickly obtain a rough but accurate solution for initializing the source term. Next, during the middle phases of training, we trained for 20000 iterations using the Adam optimizer, with exponential learning rate decay (gamma = 0.8). The sample set of collocation points had a density of 128 per unit area, and the batch size used was 512. During this phase $f_0$ was recomputed once every 1000 iterations. For the second middle phase, we trained using Adam for another 20000 iterations, only updating $f_0$ every 10000 iterations. Finally, for some improvement in accuracy, we trained for a fourth phase using a low-intensity LBFGS optimizer, a sample set density of 128 per unit area, and a batch size of 2048, and trained for 48 iterations. During this final phase, $f_0$ was recomputed every 8 iterations. During all three phases of training, we set a weight on the initial and boundary conditions (we used 1000), and to enforce timewise causality and avoid zero mode failures (as noted in section 3.2), we applied time-adaptive weighting on collocation points.

## 5. Conclusions and outlook

We have compared traditional finite volume solvers with PINN solvers for elliptic (Poisson), hyperbolic (advection), and parabolic (diffusion) equations in 2D settings and evaluated their potential applications for solving kinetic equations for electrons in collisional plasmas. Our investigations found that despite several PINN solver improvements, PINNs do not compete with traditional FV solvers in accuracy and efficiency. The PINN methods that showed the most promise in our investigations were PINN time-stepping, exploiting PINN predictivity for time-dependent problems, and adaptive sampling for generating solution-refining collocation points. Some additional conclusions can be drawn:

- Special efforts are required to ensure conservation laws (MC-PINNs) and long-time accuracy for transient simulations (LSTM or moving time window).
- The predictivity of PINNs suggests they may be usefully combined with traditional methods by balancing performance tradeoffs.

In future work,

- We plan to integrate ML algorithms with traditional hybrid (kinetic-fluid) plasma solvers
- Use ML to help select different models at different scales and best closures of reduced models for efficient simulations
- Use PINNs to improve numerical algorithms for PDE with traditional methods
- Use PINNs to find parametric solutions and solve inverse problems

## Acknowledgments

NSF EPSCoR project OIA-1655280 "FTPP: Future Technologies and enabling Plasma Processes."

## References

[1] V Kolobov, F Deluzet, Adaptive Kinetic-Fluid Models for Plasma Simulations on Modern Computer Systems - Frontiers in Physics, 2019 https://doi.org/10.3389/fphy.2019.00078
[2] Shi Jin, Asymptotic-preserving schemes for multiscale physical problems, Acta Numerica 31 (2022) 415
[3] A.J. Crilly, B. Duhig, N. Bouziani, Learning closure relations using differentiable programming: An example in radiation transport, J. Quantitative Spectroscopy and Radiative Transfer 318 (2024) 108941
[4] V Kolobov, R Arslanbekov, and D Levko, Boltzmann-Fokker-Planck kinetic solver with adaptive mesh in phase space, AIP Conference Proceedings 2132, 060011 (2019)
[5] J. Piasecki and E. Wajnryb, Long-Time Behavior of the Lorentz Electron Gas in a Constant, Uniform Electric Field, Journal of Statistical Physics, Vol. 21 (1979) 549
[6] S Gordier et al, Numerical method for Vlasov-Lorentz models, ESAIM: Proc. 10 (2001) 201
[7] M Yousfi, P Segur and Vassiliadis, Solution of the Boltzmann equation with ionization and attachment, J. Phys. D: Appl. Phys. l8 (1985) 359
[8] V N Koterov and A V Shcheprov, Nonlinear quasi-diffusion equations for calculating the electron energy spectrum in the non-uniform field of arbitrary strength, Sov J Plasma Phys 17 (1991) 857
[9] V I Kolobov, R R Arslanbekov, and H Liang, Peculiarities of charged particle kinetics in spherical plasma, J. Phys.: Conf. Ser. 2742 (2024) 012016
[10] D Ferry, *Semiconductor Transport*, Taylor & Francis (2000)
[11] V I Kolobov, J A Guzman and R R Arslanbekov, A self-consistent hybrid model of kinetic striations in low-current argon discharges, Plasma Sources Sci. Technol. 31 (2022) 035020
[12] A. Daw, J. Bu, S. Wang, P. Perdikaris, and A. Karpatne. Mitigating propagation failures in physics-informed neural networks using retain-resample-release (R3) sampling. arXiv preprint arXiv:2207.02338, 2022.
[13] https://developer.nvidia.com/modulus
[14] T.G. Grossmann, U.J. Komorowska, J. Latz, and C.-B. Schönlieb. Can physics-informed neural networks beat the finite element method? IMA Journal of Applied Mathematics, 89 (2024) 143
[15] K. Prantikos, S. Chatzidakis, L.H. Tsoukalas, and A. Heifetz. Physics-informed neural network with transfer learning (TL-PINN) based on domain similarity measure for prediction of nuclear reactor transients. Scientific Reports 13 (2023) 16840
[16] C. L. Wight and J. Zhao. Solving Allen-Cahn and Cahn-Hilliard equations using adaptive physics-informed neural networks, Commun. Comput. Phys. 29 (2021) 930